\documentclass[twocolumn]{revtex4}
\usepackage{epsfig}
\usepackage{amssymb}
\usepackage{amsmath}
\usepackage{amsfonts}
\usepackage{graphicx}
\usepackage{mathrsfs}
\usepackage[dvips]{color}
\usepackage{multirow}

\newcommand{\fn}{{\mathfrak{n}}}

\newcommand{\fz}{\mathfrak{z}}

\newcommand{\fK}{\mathfrak{K}}

\newcommand{\bk}{\mathbf{k}}

\newcommand{\bM}{\mathbf{M}}

\newcommand{\cF}{\mathcal{F}}

\newcommand{\cO}{\mathcal{O}}
\newcommand{\cP}{\mathcal{P}}

\newcommand{\cS}{\mathcal{S}}
\newcommand{\cT}{\mathcal{T}}

\newcommand{\be}{\begin{equation}}
\newcommand{\ee}{\end{equation}}
\newcommand{\bea}{\begin{eqnarray}}
\newcommand{\eea}{\end{eqnarray}}
\newcommand{\nn}{\nonumber}

\newcommand{\ed}{\end{document}}

\newcommand{\rx}{{\mbox{\small${\rm X}$}}}
\newcommand{\ry}{{\mbox{\small ${\rm Y}$}}}
\newcommand{\rz}{{\mbox{\small${\rm Z}$}}}

\newcommand{\bi}{\begin{itemize}}
\newcommand{\ei}{\end{itemize}}

\newcommand{\bce}{\begin{center}}
\newcommand{\ece}{\end{center}}

\newcommand{\sE}{\mathscr{E}}

\newcommand{\sG}{\mathscr{G}}

\newcommand{\sL}{\mathscr{L}}

\newcommand{\RE}{{\rm Re}}
\newcommand{\IM}{{\rm Im}}

\begin{document}

\title{Blowing up Light: A nonlinear amplification scheme for electromagnetic waves}

\author{Ali~Mostafazadeh$^{1,2}$, Hamed Ghaemi-Dizicheh$^2$, and Sasan Hajizadeh$^2$\\[6pt]
Departments of Mathematics$^1$ and Physics$^2$, Ko\c{c} University,\\ 34450 Sar{\i}yer,
Istanbul, Turkey}

\begin{abstract}

We use blow-up solutions of nonlinear Helmholtz equations to
introduce a nonlinear resonance effect that is capable of amplifying
electromagnetic waves of particular intensity. To achieve this, we
propose a scattering setup consisting of a Kerr slab with a negative
(defocusing) Kerr constant placed to the left of a linear slab in
such a way that a left-incident coherent TE wave with a specific
incidence angle and intensity realizes a blow-up solution of the
corresponding Helmholtz equation whenever its wavenumber $k$ takes a
certain critical value, $k_\star$. For $k=k_\star$, the solution
blows up at the right-hand boundary of the Kerr slab. For
$k<k_\star$, the setup defines a scattering system with a
transmission coefficient that diverges as $(k-k_\star)^{-4}$ for
$k\to k_\star$. By tuning the distance between the slabs we can use
this setup to amplify coherent waves with a wavelength in an
extremely narrow spectral band.  For nearby wavelengths the setup
serves as a filter. Our analysis makes use of a nonlinear
generalization of the transfer matrix of the scattering theory as
well as properties of unidirectionally invisible potentials.

\end{abstract}

\maketitle


One of the remarkable properties of nonlinear differential equations
is that their initial-value problem may not admit a global solution
even if their coefficient functions are smooth. This means that the
solution $\psi(x)$ exists in the vicinity of the initial value $x_0$
of $x$, but blows up at some $x_\star>x_0$. These so-called blow-up
solutions of nonlinear differential equations have been extensively
studied by mathematicians for decades
\cite{wong,glassey,cafarelli,ogawa}, but their physical realizations
and possible applications have not been fully explored. The purpose
of the present article is to outline a concrete physical application
of the blow-up solutions which allows for their realization in a
scattering setup and forms the basis of a nonlinear amplification
scheme for electromagnetic waves.

Consider the time-independent nonlinear Schr\"odinger equation,
    \be
    -\psi''(x)+\chi(x)\left[\zeta+\gamma |\psi(x)|^2\right] \psi(x)=\fK^2\psi(x),
        \label{NLSE0}
        \ee
where 
    \[\chi(x):=\left\{\begin{array}{ccc}
    1 & {\rm for} & x\in[0,1],\\
    0 & {\rm for} & x\notin[0,1],\end{array}\right.\]
$\zeta$, $\gamma$, and $\fK$ are real parameters, and $\fK>0$.
Suppose that
    \begin{align}
    &\gamma>0,  &&\fK^2>\zeta,
    \label{condi-1}
    \end{align}
and let  \begin{align}
    &A:=\sqrt{2(\fK^2-\zeta)/\gamma},
    &&x_\star:=\frac{\pi}{2A\sqrt{2\gamma}}.
    \label{star-def-1}
    \end{align}
Then it is easy to check that for every phase angle $\varphi$, the
function
    \be
    \psi_\star(x):=A\, e^{i\varphi}
    \sec\mbox{\large$\big[\frac{\pi}{4}\big(\frac{x}{x_\star}+
    1\big)\big]$},
    \label{sol}
    \ee
is a solution of (\ref{NLSE0}) in $[0,1]$ provided that $x_\star>1$.
According to (\ref{sol}), $\psi_\star(x)$ blows-up at $x=x_\star$.
Therefore, it defines a blow-up solution of (\ref{NLSE0}) in $[0,1]$
whenever $x_\star\leq 1$.

If $x_\star>1$, which means
    \be
    A<A_\star:=\frac{\pi}{2\sqrt{2\gamma}},
    \label{condi-A}
    \ee
we can extend (\ref{sol}) to the whole real axis to obtain the
following global solution of (\ref{NLSE0}).
    \be
    \psi_\star(x):=\left\{\begin{array}{ccc}
    c_+ e^{i\fK x}+c_- e^{-i\fK x} &{\rm for} & x<0,\\[3pt]
    Ae^{i\varphi}\sec\left[\frac{\pi}{4}\left( x/x_\star+1\right)\right]
    &{\rm for} & x\in[0,1],\\[3pt]
    d_+ e^{i\fK x}+d_- e^{-i\fK x} &{\rm for} & x>1,
    \end{array}\right.
    \label{sol2}
    \ee
where
    \bea
    c_\pm&:=&\frac{A e^{i\varphi}}{\sqrt 2}\left[1\mp\frac{i\pi(1-\epsilon)}{4\fK}\right],
    \label{cpm}\\
    d_\pm&:=&\frac{A\, e^{i(\varphi\mp\fK)}\cos(\pi\epsilon/4)}{2
    \sin^2(\pi\epsilon/4)}\left[\tan(\mbox{\large$\frac{\pi\epsilon}{4}$})\mp
    \frac{\pi i (1-\epsilon)}{4\fK}\right],~~
    \label{dpm}
    \eea
    \bea
    \epsilon&:=&1-A/A_\star=1-x_\star^{-1}.
    \label{epsilon-def}
    \eea
so that $\psi_\star(x)$ is continuous and differentiable at $x=0$
and $x=1$.

In view of (\ref{epsilon-def}),
    \begin{align}
    &A=A_\star(1-\epsilon),
    && x_\star=\frac{1}{1-\epsilon}=1+\epsilon+\cO(\epsilon^{2}),
         \label{A-x-exp}
         \end{align}
where $\cO(\epsilon^n)$ stands for terms of order $n$ and higher in
powers of $\epsilon$. Substituting (\ref{A-x-exp}) in the first
equation in (\ref{star-def-1}), we have
    \be
    \fK=\fK_\star\sqrt{1-\frac{\pi^2\epsilon(2-\epsilon)}{16\fK_\star^2}}=
    \fK_\star-\frac{\pi^2\epsilon}{16\fK_\star}+\cO(\epsilon^2),
    \label{K-exp}
    \ee
where
    \[\fK_\star:=\sqrt{\zeta+\frac{\pi^2}{16}}.\]
Expanding the right-hand side  of (\ref{cpm}) and (\ref{dpm}) and
making use of (\ref{A-x-exp}) and (\ref{K-exp}), we find
    \bea
    c_\pm&=&\frac{A e^{i\varphi}}{\sqrt 2}
    \left(1\mp\frac{i\pi}{4\fK_\star}\right)+\cO(\epsilon),
    \label{c-exp}\\
    d_\pm&=& \mp \frac{2 i A e^{i(\varphi\mp\fK_\star)}}{\pi\fK_\star\,\epsilon^2}+
    \cO(\epsilon^{-1}).
    \label{d-exp}
    \eea
According to (\ref{epsilon-def}) -- (\ref{d-exp}), if we arrange
that $ \fK \to \fK_\star $, so that $\epsilon\to 0$, $c_\pm$ tend to
finite values while $d_\pm$ diverge quadratically.

\begin{figure}
    \begin{center}
    \includegraphics[scale=.35,clip]{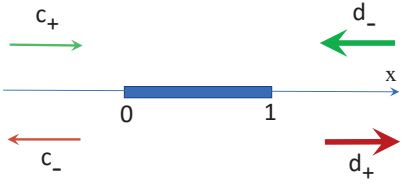}
    \caption{(Color online) Schematic view of a Kerr nonlinearity
    confined to a finite interval, i.e., $[0,1]$, on the $x$-axis.
    $c_+$ and $d_-$ are respectively the complex amplitudes of a pair
    of left- and right-incident waves. These are scattered into the
    outgoing waves of amplitude $c_-$ and $d_+$. The thicker arrows
    represent higher intensity waves.
    \label{fig01}}
    \end{center}
    \end{figure}

The solution~(\ref{sol2}) corresponds to a situation where a pair of
right- and left-going incident plane waves, $c_+e^{i(\fK x-\omega
t)}$ and $d_-e^{-i(\fK x+\omega t)}$, are  scattered by a  confined
nonlinearity \cite{prl-2013} into the left- and right-going
scattered waves: $c_-e^{-i(\fK x+\omega t)}$ and $d_+e^{i(\fK
x-\omega t)}$, as depicted in Fig.~\ref{fig01}.
For $\epsilon\approx 0$, the right-incident wave that is sent from
$x=+\infty$ and the right-going scattered wave that reaches
$x=+\infty$ have a much larger amplitude than the left-incident wave
and the left-going scattered wave. This means that the Kerr
nonlinearity acts as a filter for the high-intensity incident wave
from the right, namely $d_-e^{-i(\fK x+\omega t)}$, provided that we
inject it from the left by the much lower intensity wave
$c_+e^{i(\fK x-\omega t)}$. This is actually a curious observation,
but is not what we wish to accomplish. Our goal is to explore the
possibility of introducing a genuine scattering setup in which a
blow-up solution is realized by an incident plane wave that is sent
only from the left or the right. The singular nature of the solution
would then imply a substantial amplification of the transmitted
wave. This signifies a nonlinear amplification scheme that we intend
to utilize in optics.

As a first step in this direction, we employ the equivalence of the
nonlinear Schr\"odinger equation (\ref{NLSE0}) with the Helmholtz
equation describing the interaction of a transverse electric (TE)
wave with a Kerr slab
\cite{marburger,yeh,chen,peterson-1990,prl-2013,jo-2017,pla-2017}.

Consider an infinite planar Kerr slab of thickness $L$ that is
placed in a nonmagnetic homogeneous linear medium filling the space
outside the slab and having a real refractive index $n_0\geq 1$ .
Suppose that we choose a cartesian coordinate system
$\{(\rx,\ry,\rz)\}$ in which the slab occupies the space bounded by
the planes $\rz=0$ and $\rz=L$, and $\widehat\varepsilon_l$ denotes
the linear relative permittivity of the slab. Then it is easy to
show \cite{jo-2017} that the electric field for a time-harmonic TE
wave interacting with this system has the form:
$\exp[i(n_0k\sin\theta\,\rx-\omega t)]\sE(\rz) \hat
e_{\mbox{\scriptsize Y}}$, where $k$ is the wavenumber, $\theta$ is
the incidence angle of the wave, $\omega:=ck$, $c$ is the speed of
light in vacuum, $\sE(\rz)$ is the complex amplitude of the electric
field, and $\hat e_j$ is the unit vector pointing along the $j$-axis
for $j=\rx,\ry,\rz$. Using Maxwell's equation, we can show that
$\sE(\rz)$ satisfies the Helmholtz equation,
    \be
    \sE''(\rz)+k^2\,\widehat{\mbox{\large$\varepsilon$}}
    (\rz,\sE)\,\sE(\rz)=0,
    \label{helmholtz-eq}
    \ee
where
    \be
    \widehat{\mbox{\large$\varepsilon$}}(\rz,\sE)
    :=\left\{\begin{array}{ccc}
    \widehat\varepsilon_l-\sin^2\theta+\sigma|\sE(\rz)|^2
    &{\rm for} & \rz\in[0,L],\\[6pt]
    n_0^2-\sin^2\theta&{\rm for} & \rz\notin[0,L],\end{array}\right.
    \label{permi}
    \ee
and $\sigma$ is the nonlinearity (Kerr) coefficient. Imposing the
electromagnetic interface conditions at the faces of  the slab
\cite{jackson}, we find that $\sE$ and $\sE'$ must be continuous at
$\rz=0$ and $\rz=L$.

In terms of the scaled parameters:
    \begin{align}
    &x:=\frac{\rz}{L},
    &&\fK:=kL\sqrt{n_0^2-\sin^2\theta},
    \label{new1}\\
    &\zeta:=k^2L^2(n_0^2-\widehat\varepsilon_l),
    &&\gamma:=-k^2L^2\sigma,
    \label{new2}
    \end{align}
the Helmholtz equation (\ref{helmholtz-eq}) takes the form of the
nonlinear Schr\"odinger equation (\ref{NLSE0}) provided that we set
$\psi(x):=\sE(Lx)$. In view of this relation and
Eqs.~(\ref{condi-1}) and (\ref{new2}), (\ref{sol}) gives a blow-up
solution of (\ref{helmholtz-eq}), if
    \begin{align}
    &\sigma<0, &&  \widehat\varepsilon_l>\sin^2\theta.
    \label{condi-2}
    \end{align}
Therefore we need a Kerr slab with negative (defocusing) Kerr
coefficient. Refs.~\cite{li,zhang-2017} study particular examples of
Kerr media with negative Kerr coefficient. See alse \cite{neiral}.


According to (\ref{star-def-1}), (\ref{condi-A}), (\ref{A-x-exp}),
(\ref{new1}), and (\ref{new2}),
    \begin{align}
    &\begin{aligned}
    A= \sqrt{\frac{2(\widehat\varepsilon_l-\sin^2\theta)}{-\sigma}},
    \end{aligned}
    \label{A-k=2}\\
    &\begin{aligned}
    &k=k_\star(1-\epsilon),~~~  &&
    \fK =\tilde\fK_\star(1-\epsilon),
    \end{aligned}
    \label{kk=kk}
    \end{align}
where 
    \begin{align}
    & k_\star:=\frac{\pi}{4L\sqrt{\widehat\varepsilon_l-\sin^2\theta}},
    && \tilde\fK_\star:=
    \frac{\pi}{4}\sqrt{\frac{n_0^2-\sin^2\theta}{\widehat\varepsilon_l-\sin^2\theta}}.
    \label{k-star-def2}
    \end{align}
Substituting (\ref{A-k=2}) in (\ref{cpm}) and (\ref{dpm}), we find
    \bea
    c_\pm&=&\frac{A\, e^{i\varphi}}{\sqrt 2}\left(1\mp\frac{i\pi}{4\tilde\fK_\star}\right),
    \label{c-exp-2}\\
    d_\pm&=& \mp \frac{2 i A\, e^{i(\varphi\mp \tilde\fK_\star)}}{\pi\tilde\fK_\star\,\epsilon^2}+
    \cO(\epsilon^{-1}).
    \label{d-exp-2}
    \eea
Equations~(\ref{kk=kk}) and (\ref{d-exp-2}) show that $d_\pm$ have a
quadratic divergence at $k=k_\star$.

Next, we return to the main missing step towards using blow-up
solutions for the purpose of amplifying waves, namely devising a
genuine scattering system whose transmission coefficient diverges
for certain values of the intensity and wavenumber of the incident
wave. To do this, first we recall the basic framework for scattering
by confined nonlinearities and outline a nonlinear generalization of
the transfer matrix of linear scattering theory which proves to be a
useful tool for performing the necessary calculations.

Consider the wave equation
    \be
    -\psi''(x)+[v(x)+\cF(x,\psi)]\psi(x)=\fK^2\psi(x),
    \label{NLSE}
    \ee
where $v(x)$ and $\cF(x,\psi)$ are functions representing the linear
and nonlinear interactions of a physical system, respectively.
Suppose that for $x\to\pm\infty$ these functions decay to zero at
such a rate that the global solutions of (\ref{NLSE}) tend to plane
waves at spatial infinities, i.e.,
    \bea
    \psi(x)&\to& A_- e^{i\fK x}+B_- e^{-i\fK x}~~~{\rm for}~~~x\to-\infty,
    \label{ini-m}\\
    \psi(x)&\to& A_+ e^{i\fK x}+B_+ e^{-i\fK x}~~~{\rm for}~~~x\to\infty,
    \label{ini-p}
    \eea
where $A_\pm$ and $B_\pm$ are complex coefficients.

The scattering solutions $\psi_{\rm l/r}$ of (\ref{NLSE}) that
respectively correspond to a left/right-incident wave of complex
amplitude $A^{\rm l/r}$ satisfy the asymptotic boundary conditions:
    \bea
    \psi_{\rm l}(x)&\to&\left\{\begin{array}{ccc}
    A^{\rm l} \left(e^{i\fK x}+R^{\rm l}e^{-i\fK x}\right) & {\rm for} & x\to-\infty,\\
    A^{\rm l} T^{\rm l}e^{i\fK x} & {\rm for} & x\to+\infty,\end{array}\right.~~~
    \label{psi-left}\\
    \psi_{\rm r}(x)&\to&\left\{\begin{array}{ccc}
    A^{\rm r}T^{\rm r}e^{-i\fK x}& {\rm for} & x\to-\infty,\\
    A^{\rm r} \left(e^{-i\fK x}+R^{\rm r}e^{i\fK x}\right)  &
    {\rm for} & x\to+\infty,\end{array}\right.
    \label{psi-right}
    \eea
where $R^{\rm l/r}$ and $T^{\rm l/r}$ are respectively the
left/right reflection and transmission amplitudes
\cite{muga,bookchapter}. In the absence of nonlinearity these are
complex-valued functions of $\fK$, but in general they depend on
both $\fK$ and $A^{\rm l/r}$, \cite{prl-2013}.

The scattering problem defined by (\ref{NLSE}) admits a
transfer-matrix formulation \cite{p148}. For a solution specified by
its asymptotic form at $x=-\infty$, equivalently the coefficients
$A_-$ and $B_-$ entering (\ref{ini-m}), we can identify the transfer
matrix with a $2\times 2$ matrix $\bM$ satisfying
    \be
    \left[\begin{array}{cc}
    A_+\\B_+\end{array}\right]=\bM\left[\begin{array}{cc}
    A_-\\B_-\end{array}\right].
    \label{M=}
    \ee
For the well-known linear interactions where $\cF(x,\psi)=0$, this
equation defines $\bM$ as a unique $2\times 2$ matrix that does not
depend on $A_-$ and $B_-$. In this case the entries of $\bM$ are
functions of $\fK$ and its determinant equals unity. In the presence
of nonlinearities, $\det\bM$ may deviate from unity, and the entries
of $\bM$ depend also on $A_-$ and $B_-$. In this case, (\ref{M=})
does not determine $\bM$ in a unique manner, but we can use
(\ref{psi-left}) and (\ref{psi-right}) to relate any choice of $\bM$
satisfying (\ref{M=}) to the reflection and transmission amplitudes
in the form
    \begin{align}
    &R^{\rm l}=-M_{21}^{\rm l}/M_{22}^{\rm l},
    &&T^{\rm l}=\det\bM^{\rm l}/M_{22}^{\rm l},
    \label{RT-left}\\
    &R^{\rm r}=M_{12}^{\rm r}/M_{22}^{\rm r},
    &&T^{\rm r}=1/M_{22}^{\rm r},
    \label{RT-right}
    \end{align}
where $M_{ij}^{\rm l/r}$ are the entries of
    \begin{align}
    &\bM^{l}:=\bM(A^{l},A^{l} R^{l}),
    &&\bM^{r}:=\bM(0,A^{r} T^{r}).
    \label{M-left-right}
    \end{align}
In practice, we can determine $\bM(A_-,B_-)$ for arbitrary choices
of $A_-$ and $B_-$ by solving the initial-value problem defined by
(\ref{NLSE}) and (\ref{ini-m}) and using (\ref{M=}).
Eqs.~(\ref{RT-left}) and (\ref{RT-right}) hold for any
$\bM(A_-,B_-)$ that we obtain in this way. In view of
(\ref{M-left-right}), these provide four complex equations for the
four unknowns $R^{{l}/{r}}$ and $T^{{l}/{r}}$.

An important advantage of the above nonlinear transfer-matrix
formulation of scattering theory is that the transfer matrix $\bM$
shares the composition property of its linear analog
\cite{prl-2009,sanchez,ap-2015,bookchapter}. To explain what we mean
by this property, suppose that there is a real number $x_0$ such
that we can decompose the interaction term $v(x)+\cF(x,\psi)$ in
(\ref{NLSE}) into the sum of two separate parts, i.e.,
    \[v(x)+\cF(x,\psi)=\sum_{j=1}^2\big[v_j(x)+\cF_j(x,\psi)\big],\]
where $v_1(x)+\cF_1(x,\psi)=0$ for $x>x_0$ and
$v_2(x)+\cF_2(x,\psi)=0$ for $x<x_0$. Then we can use (\ref{M=}) to
show that the transfer matrix $\bM^{(j)}$ associated with the
interaction $v_j(x)+\cF_j(x,\psi)$ satisfies
    \be
    \bM^{(2)}(A_0,B_0)\bM^{(1)}(A_-,B_-)=\bM(A_-,B_-),
    \label{compose}
    \ee
where
    \[\mbox{\small$\left[\begin{array}{c}A_0\\B_0\end{array}\right]$}:=\bM^{(1)}(A_-,B_-)
\mbox{\small$\left[\begin{array}{c}A_-\\B_-\end{array}\right]$}.\]

We refer to (\ref{compose}) as the composition property of nonlinear
transfer matrices, and abbreviate it as
$\bM^{(2)}\circ\bM^{(1)}=\bM$. For example, consider a case where
$\cF(x,\psi)=\gamma|\psi(x)|^2\chi(x)$, $v(x)=v_1(x)+v_2(x)$,
    \bea
    v_1(x)&=&\left\{\begin{array}{ccc}
    \zeta &{\rm for}~~x\in[0,1],\\
    0 &{\rm otherwise},\end{array}\right.\\
    v_2(x)&=&\left\{\begin{array}{ccc}
    \fz(x) &{\rm for}~~x\in[a,a+\ell],\\
    0 &{\rm otherwise},\end{array}\right.
    \eea
$\zeta$, $a$, and $\ell$ are real parameters, $a>1$, $\ell>0$, and
$\fz(x)$ is a real- or complex-valued function. Then we can express
the transfer matrix for (\ref{NLSE}) as
    \be
    \bM=\bM^{(2)}\circ\bM^{(1)}
    \label{compose2}
    \ee
where  $\bM^{(1)}$ and $\bM^{(2)}$ are respectively the transfer
matrices for the interactions $v_1(x)+\gamma|\psi(x)|^2$ and
$v_2(x)$. In particular $\bM^{(2)}$ is uniquely determined by the
reflection and transmission amplitudes of the finite-range potential
$v_2(x)$. Denoting these by $R^{{l}/{r}}_2$ and $T^{{l}/{r}}_2$, and
recalling that scattering potentials enjoy transmission reciprocity
\cite{ahmed-2001,prl-2009,bookchapter}, so that
$T_2^{l}=T_2^{r}=:T_2$, we have
    \be
    \begin{aligned}
    &M^{(2)}_{11}=T_2-R_2^{l} R_2^{r}/T_2,~~~
    &&M^{(2)}_{12}= R_2^{r}/T_2,\\
    &M^{(2)}_{21}=- R_2^{l}/T_2,
    &&M_{22}^{(2)}= 1/T_2.
    \end{aligned}
    \label{Mij=RT}
    \ee

\begin{figure}
    \begin{center}
    \includegraphics[scale=.4,clip]{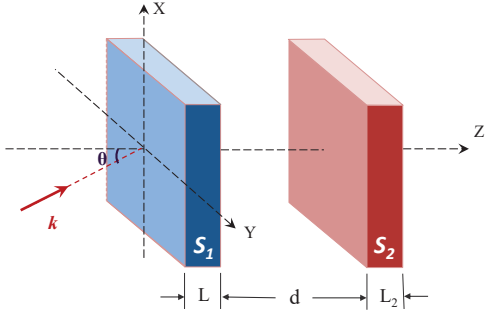}
    \caption{(Color online) Schematic view of a scattering system consisting of a Kerr slab
    $\cS_1$ and a linear slab $\cS_2$. $L$ and $L_2$ are respectively the thickness of $\cS_1$ and
    $\cS_2$, and $d$ is their distance. $\bk$ represents the wavevector for an incoming TE wave with
    incidence angle $\theta$.
    \label{fig1}}
    \end{center}
    \end{figure}

The scattering setup we outline in the preceding paragraph admits an
optical realization involving a homogeneous Kerr slab $\cS_1$ and a
nonmagnetic linear slab $\cS_2$ that is placed to the right of
$\cS_1$, as depicted in  Fig.~\ref{fig1}.
Again we assume that the space outside the slabs is filled with a
homogeneous dielectric medium with a real refractive index $n_0\geq
1$, and consider the scattering of the TE waves. Then $a,\ell$, and
$\fz(x)$ are related to the distance $d$ between the slabs,  and the
thickness $L_2$ and refractive index $\fn_2(x)$ of $\cS_2$ according
to $a=d/L+1$, $\ell=L_2/L$, and
    \begin{align}
    \fz(x)=k^2L^2[n_0^2-\fn_2(x)^2].
    \end{align}
The Helmholtz equation describing the interaction of the TE waves
with this system admits a blow-up solution, if (\ref{condi-2})
holds. The role of the linear slab is to realize a (near) blow-up
solution that fulfills the asymptotic boundary
condition~(\ref{psi-left}). In other words, we wish to construct a
solution of the form
    \be
    \psi_\star(x):=\left\{\!\!\begin{array}{ccc}
    c_+ e^{i\fK x}+c_- e^{-i\fK x} &{\rm for} & x<0,\\[6pt]
    A e^{i\varphi}\sec\!\left\{\frac{\pi}{4}\!
    \left[(1-\epsilon)x+1\!\right]\right\} &{\rm for} & x\in[0,1],\\[6pt]
    d_+ e^{i\fK x}+d_- e^{-i\fK x} &{\rm for} & x\in(1,a),\\[6pt]
    \phi(x) &{\rm for} & x\in[a,a+\ell],\\[6pt]
    c_+ T^{l}   e^{i\fK x}&{\rm for} & x>a+\ell,
    \end{array}\right.
    \label{sol21}
    \ee
where $c_\pm$ and $d_\pm$ are given by (\ref{cpm}) and (\ref{dpm}),
and $\phi(x)$ is the solution  of
$-\phi''(x)+\fz(x)\phi(x)=\fK^2\phi(x)$ in $[a,a+\ell]$ that ensures
the continuity and differentiability of $\psi(x)$ at $x=a$ and
$x=a+\ell$.
    \begin{figure}
    \begin{center}
    \includegraphics[scale=.3,clip]{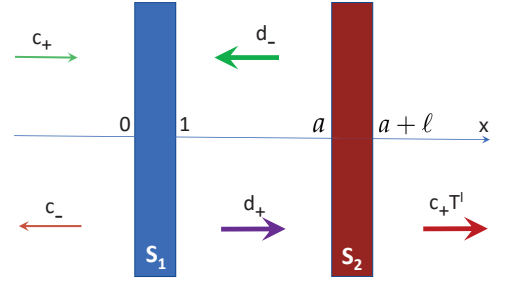}
    \caption{(Color online) Graphical demonstration of the structure of
    the solution~(\ref{sol21}). $S_1$ and $S_2$ are respectively the
    nonlinear (Kerr) and linear slabs, which in terms of $x:=\rz/L$,
    $a=d/L+1$, and $\ell=L_2/L$ correspond to intervals $[0,1]$ and
    $[a,a+\ell]$ on the $x$-axis. $c_+$, $c_-$, and $c_+T^l$
    denote the complex amplitude of the incident, reflected, and
    transmitted waves. Again the thicker arrows represent higher
    intensity waves.
    \label{fig3}}
    \end{center}
    \end{figure}

According to (\ref{M=}), (\ref{compose2}), and (\ref{Mij=RT}),
    \bea
    \left[\!\begin{array}{c}
    c_+ T^{l}\\
    0\end{array}\!\right]&=&\bM\left[\!\begin{array}{c}
    c_+\\
    c_-\end{array}\!\right]=\bM^{(2)}\!\circ \bM^{(1)}\!\left[\begin{array}{c}
    c_+\\
    c_-\end{array}\!\right]=\bM^{(2)}\!\left[\begin{array}{c}
    d_+\\
    d_-\end{array}\!\right]\nn\\
    &=&\frac{1}{T_2}
    \left[\begin{array}{c}
    (T_2^2-R_2^{l} R_2^{r})d_++R_2^{r} d_-\\
    -R_2^{l} d_++d_-\end{array}\right].\nn
    \eea
This in turn implies
    \begin{align}
    &R_2^{l}=d_-/d_+,
    &&T^{l}=T_2 d_+/c_+.
    \end{align}
Substituting (\ref{cpm}) and (\ref{dpm}) in these equations and
making use of (\ref{A-x-exp}) and (\ref{A-k=2}), we find that, for
$k=k_\star(1-\epsilon)$,
    \bea
    R_2^{l}&=&e^{2i\tilde\fK_\star(1-\epsilon)}\left[
    \frac{4\tilde\fK_\star\tan(\pi\epsilon/4)+i\pi}{4\tilde\fK_\star\tan(\pi\epsilon/4)-i\pi}\right]
    \label{R2=}\\
    &=&-\exp\left\{2i\tilde\fK_\star\left[1-2\epsilon+\cO(\epsilon^2)\right]\right\},
    \nn
    \\[6pt]
    T^{l}&=&\frac{T_2 \,e^{-i\tilde\fK_\star(1-\epsilon)}}{\sqrt 2\sin^2(\pi\epsilon/4)}
    \left[\frac{4\tilde\fK_\star\sin(\pi\epsilon/4)-i\pi\cos(\pi\epsilon/4)}{4\tilde\fK_\star-i\pi}
    \right]~~~~\nn\\
    &=&T_2\left[\frac{8\sqrt 2 e^{-i\tilde\fK_\star}}{\pi(\pi+4i\tilde\fK_\star)\epsilon^2}+\cO(\epsilon^{-1})\right].
    \label{T2=}
    \eea

Equation (\ref{R2=}) shows that $|R_2^{l}|=1$. If $\fz(x)$ is
real-valued, we can use the unitarity condition,
$|R_2^{{l}/{r}}|^2+|T_2|^2=1$, to infer that $T_2=0$. But it is
well-known that the transmission amplitude for a scattering
potential never vanishes \cite{bookchapter}. This means that in
order to realize the near-blow-up scattering solution (\ref{sol21}),
we must employ a linear medium with a complex refractive index, so
that $\fz(x)$ takes complex values and the unitarity relation need
not hold.  We also demand that for $\fK\approx\fK_\star$ (which
means $k\approx k_\star$), the transmission coefficient of this slab
is not too small. This implies that $|R_2^{{l}}|^2+|T_2|^2>1$. It is
not difficult to see that this inequality can be satisfied only if
$\cS_2$ includes gain regions. We give a rigorous proof of this
statement in the appendix.

Ref.~\cite{pra-2014a} provides an explicit construction of
finite-range potentials with any given reflection and transmission
amplitudes at a given wavenumber. Because we need a potential
$v_2(x)$ that has a sizable transmission amplitude and unit left
reflection coefficient, we use a unidirectionally right-invisible
potential $u(x)$ with support  $[0,\ell]$ and a unit left reflection
coefficient at $\fK=\fK_\star$. Such a potential fulfills all our
requirements except that the phase of its left reflection amplitude
may not coincide with that of (\ref{R2=}). Because translations,
$x\to x-a$, of a finite-range potential change its left reflection
amplitude according to,
   $R^{l}\to e^{2ia\fK} R^{l}$, we set
    \be
    v_2(x)=u(x-a),
    \label{v2=}
    \ee
and adjust $a$ such that the left reflection amplitude of $v_2$ at
$\fK=\fK_\star$ is given by (\ref{R2=}). This determines $a$ up to
an integer multiple of $\pi/\fK_\star$, \cite{pra-2014a}. We note
that adjusting the value of $a$ corresponds to tuning the distance
between the slabs. Moreover, because $v_2(x)$ is unidirectionally
invisible, $T_2=1$. Therefore, according to (\ref{T2=}), the
transmission amplitude of our two-slab system diverges quadratically
for $k\to k_\star$. Equivalently, its left transmission coefficient,
$|T^{l}|^2$, has a quartic divergence at this wavenumber.

Next, we examine the time-averaged nonlinear relative permitivity of
the Kerr slab in the vicinity of its left-hand boundary, i.e.,
$x=0$. According to  (\ref{c-exp-2}) and (\ref{sol21}), this is
given by
    \be
    \widehat\varepsilon_{nl}=\frac{1}{2}\sigma|\psi_\star(0)|^2=\sigma\,A^2.
    \label{nl-eps}
    \ee
Let us express this quantity in terms of the (time-averaged)
intensity  $I$ of the incident wave. To do this we write the
refractive index of the Kerr slab at $\rz=0$ in the form
$\fn_1=\sqrt{\widehat\varepsilon_l}+n_2I$ where $n_2$ is a negative
real constant. Because typically $|n_2|I\ll
\sqrt{\widehat\varepsilon_l}$ and
$\fn_1^2=\hat\varepsilon_l+\hat\varepsilon_{nl}$, Eq.~(\ref{nl-eps})
implies that
    \be
    |\sigma|\,A^2=|n_2|I \left(2\sqrt{\widehat\varepsilon_l}-|n_2|I \right)\approx
    2 \sqrt{\widehat\varepsilon_l}\, |n_2|I .
    \label{A-bound}
    \ee
Combining this relation with (\ref{condi-2}) and (\ref{A-k=2}), and
noting that $\widehat\varepsilon_{nl}\ll 1$, we find:
$0<\widehat\varepsilon_l-\sin^2\theta=|\sigma|A^2/2\ll 1$. Because
$\sin^2\theta<1$, this relation implies that the Kerr slab must be
made of a  (meta)material whose linear relative permittivity is
smaller than unity. Using such a Kerr slab we can realize the
proposed nonlinear resonance effect for a TE wave provided that its
incidence angle $\theta$ is slightly smaller than ${\rm
arcsin}(\widehat\varepsilon_l)$.
In particular for a normally incidence TE wave,  we need a
metamaterial with a negative Kerr coefficient and a nearly zero
linear permittivity \cite{neiral,kaipurath,caspani,alam}. Notice,
however, that according to (\ref{A-bound}),
$\widehat\varepsilon_l>n_2^2I^2/4$.

For a given Kerr slab $\cS_1$ with $\sigma<0$ and
$\widehat\varepsilon_l<1$, we choose the incidence angle $\theta$ of
the TE wave such that $0<{\rm
arcsin}(\widehat\varepsilon_l)-\theta\ll 1$. We can then compute the
value of $A$, $k_\star$, and $\tilde\fK_\star$ using (\ref{A-k=2})
and (\ref{k-star-def2}). Next, we choose a linear slab $\cS_2$ that
is unidirectionally right-invisible for $k=k_\star(1-\epsilon)$,
with  $0<\epsilon\ll 1$, and has a unit left-reflection coefficient
($|R_2^{l}|^2=1$) at this wavenumber. We place $\cS_2$ at a distance
$d$ to the right of $\cS_1$ such that  (\ref{R2=}) holds for
$k=k_\star(1-\epsilon)$. Finally, we prepare a left-incident TE wave
with incidence angle $\theta$, wavenumber $k= k_\star(1-\epsilon)$,
and time-averaged intensity $I$.



It is not difficult to see that the above conditions  restrict the
thickness $L$ of the Kerr slab. Let $\lambda:=2\pi/k$ be the
wavelength of the incident wave. Then (\ref{kk=kk}),
(\ref{k-star-def2}), and (\ref{nl-eps}) suggest that
    \[L=\frac{\lambda}{4A\sqrt{2 |\sigma|}}=\frac{\lambda}{4\sqrt{2 |\widehat\varepsilon_{nl}|}}.\]

\begin{figure}
    \begin{center}
    \includegraphics[scale=.45,clip]{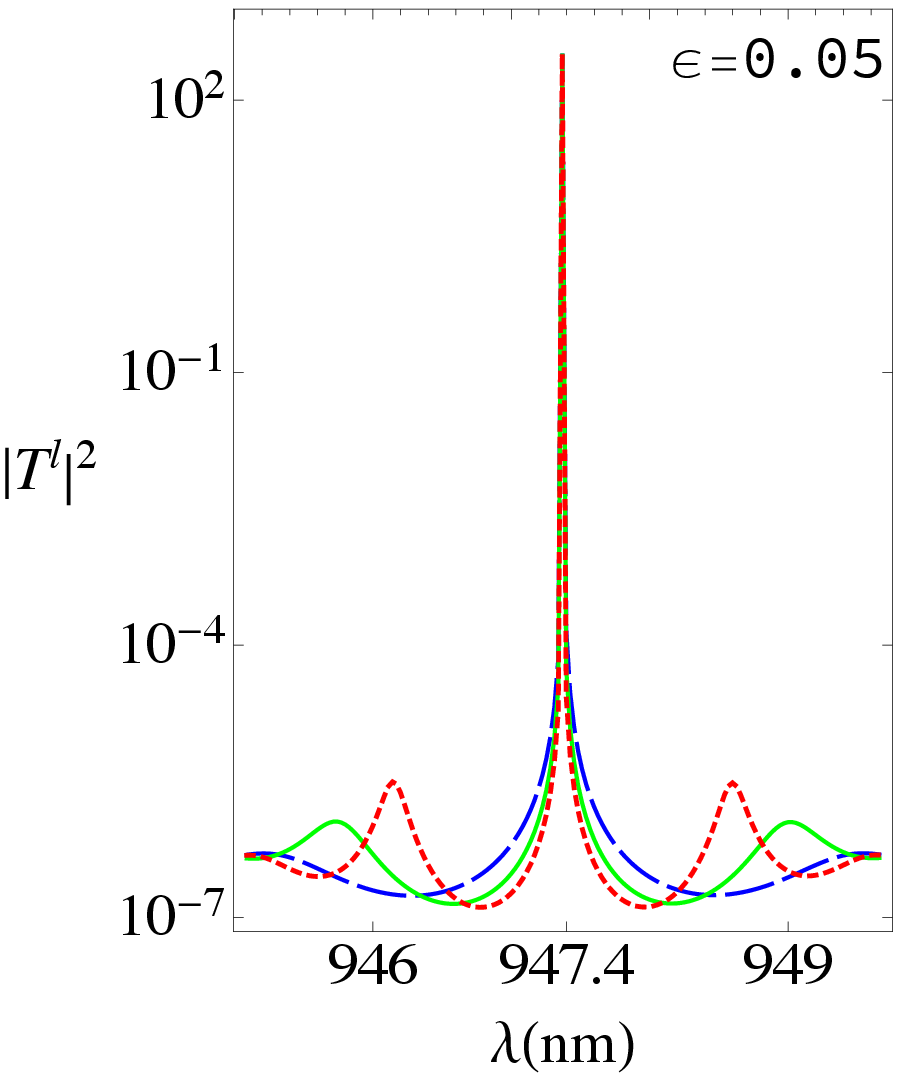}~~~~
    \includegraphics[scale=.45,clip]{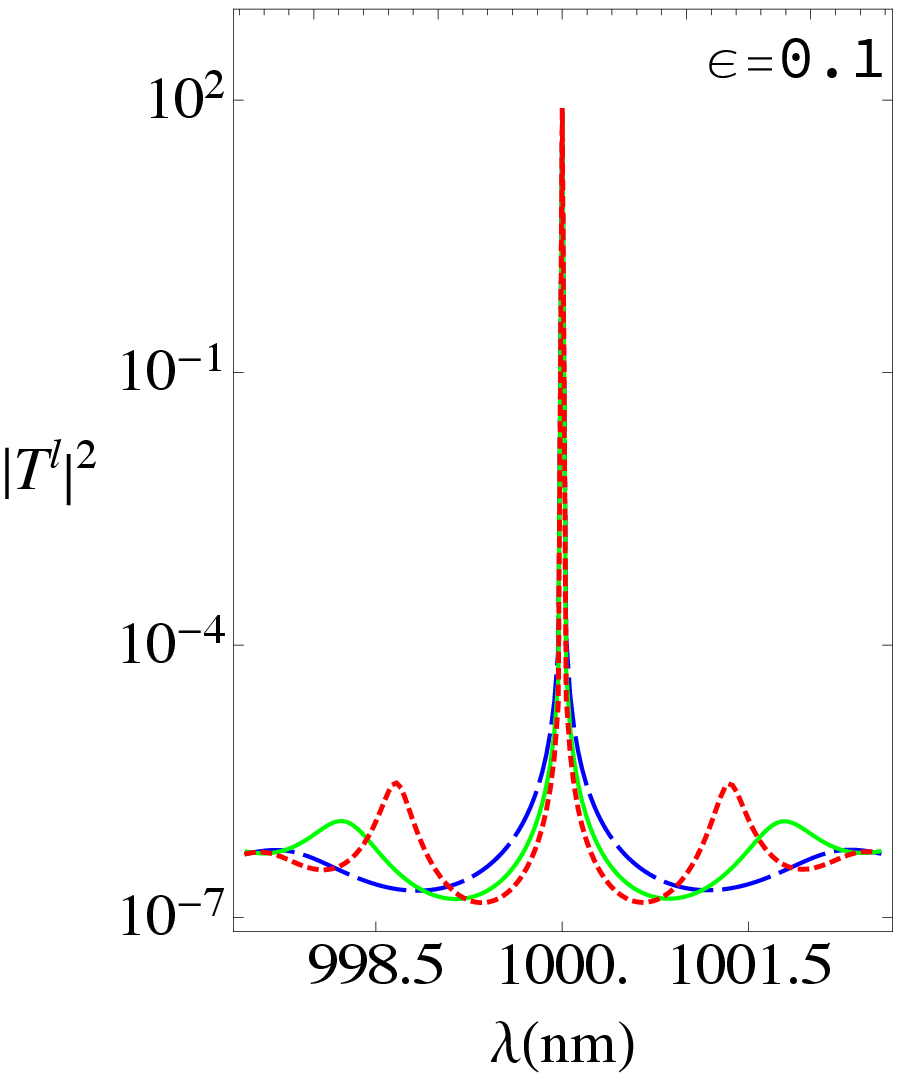}
    \caption{(Color online) Plots of the transmission coefficient
    $|T^l|^2$ as a function of the wavelength $\lambda$ when
    the optical potential describing the linear slab is given by
    (\ref{v2=}), (\ref{u=}), and (\ref{L2=}). For the graph on the left,
    $\epsilon=0.05$, $d=10.162 \mu{\rm m}$ (dashed blue curve), $100.162 \mu{\rm m}$ (solid green curve), and $200.109 \mu{\rm m}$ (dotted red curve). For the graph on the right, $\epsilon=0.10$, $d=10.471 \mu{\rm m}$ (dashed blue curve), $100.471 \mu{\rm m}$ (solid green curve), and $200.471 \mu{\rm m}$ (dotted red curve).
    The peak value of $|T^l|^2$ are respectively $324.536$ and $81.247$
    for $\epsilon=0.05$ and $0.10$. They occur for $\lambda=947.368$
    and $1000.000~{\rm
    nm}$.
    \label{fig2}}
    \end{center}
    \end{figure}

Figure~\ref{fig2} shows the  plots of the transmission coefficient
of our two-slab system for different values of $\epsilon$ and $d$.
This corresponds to the scattering of a left-incident wave with
intensity $I=1\, {\rm GW}/{\rm cm}^2$ where the Kerr slab has
thickness $L=503.115\,\mu{\rm m}$, linear relative permittivity
$\widehat\varepsilon_l=0.25$, and Kerr constant $n_2=-10^{-16}\,{\rm
cm}^2/W$, so that the incident angle of the wave is to be taken as
$\theta=29.999997^\circ$, \cite{footnote1}. Both slabs is placed in
vacuum, i.e., $n_0=1$. The linear slab is modeled using the
right-invisible optical potential (\ref{v2=}) with $u(x)$ given by
\cite{pra-2014a}:
    \be
    u(x):=\left\{\begin{array}{cc}
    \displaystyle -\frac{8\alpha \fK^2 (3-2 e^{2i\fK_0x})}{
    e^{4i\fK_0x}+\alpha(1-e^{2i\fK_0x})^2}& {\rm for}~ x\in[0,\ell],\\[12pt]
    0 &{\rm otherwise},\end{array}\right.
    \label{u=}
    \ee
where $\alpha=-10^{-4}$, $\fK_0:=\fK_\star(1-\epsilon)$,
$\ell:=L_2/L=400\pi/\fK_0$, $a=1+d/L$,
$\lambda_\star=2\pi/k_\star=900~{\rm nm}$, and
    \be
    L_2=\left\{\begin{array}{ccc}
    189.474\,\mu{\rm m}  &{\rm for}~\epsilon=0.05,\\[3pt]
    200.000 \,\mu{\rm m} &{\rm for}~\epsilon=0.10.
    \end{array}\right.
    \label{L2=}
    \ee
As expected $|T^l|^2$ has a sharp peak at
$\lambda=\lambda_0:=2\pi/k_\star(1-\epsilon)$. This is a clear
demonstration of the nonlinear resonance effect that we describe
above. Notice that for wavelengths slightly different from
$\lambda_0$ the transmission coefficient takes extremely small
values. This shows that our setup acts as a highly effective filter
for small deviations from the resonance wavelength $\lambda_0$.

\begin{figure}
    \begin{center}
    \includegraphics[scale=.45,clip]{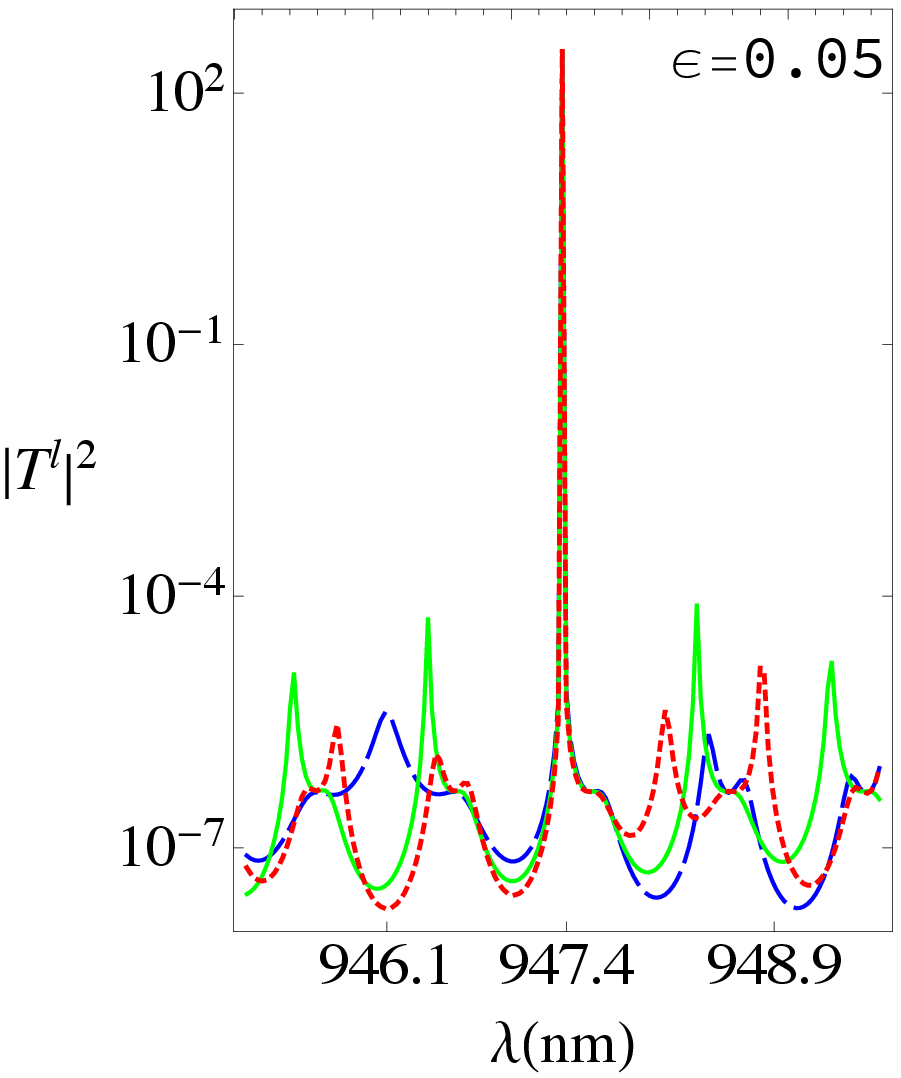}~~~~
    \includegraphics[scale=.45,clip]{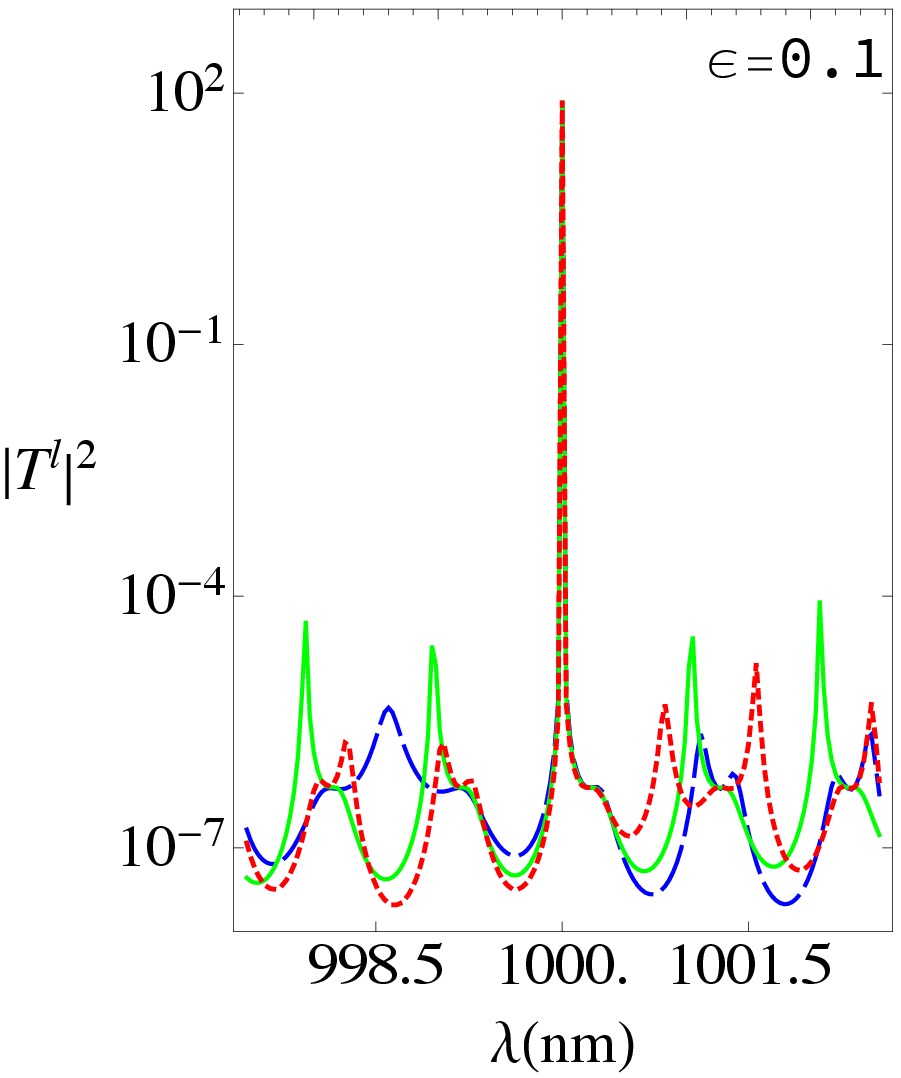}
    \caption{Plots of the transmission coefficient $|T^l|^2$ as a function of
    the wavelength $\lambda$ when the linear slab is a $\cP\cT$-symmetric bilayer determined by (\ref{v2=}), (\ref{PT-u=}), and (\ref{PT-L2=}). For the graph on the left, $\epsilon=0.05$, $d=10.162 \mu{\rm m}$ (dashed blue curve),
   $100.162 \mu{\rm m}$ (solid green curve), and
   $200.109 \mu{\rm m}$ (dotted red curve). For the graph on the right,
   $\epsilon=0.10$, $d=10.471 \mu{\rm m}$ (dashed blue curve),
   $100.471 \mu{\rm m}$ (solid green curve), and
   $200.471 \mu{\rm m}$ (dotted red curve).
   The position of the peaks and their height are the same as in Fig.~\ref{fig2}, because they are determined by the properties of the incident wave and the Kerr slab.
    \label{fig5}}
    \end{center}
    \end{figure}

The main reason for our choice of (\ref{u=}) for the function $u(x)$
is that it involves the free parameter $\alpha$ which we can tune to
set the left reflection coefficient of the linear slab to unity,
i.e., make $|R^l|^2=1$ for $\lambda=\lambda_0$. We can achieve the
same purpose using a right-invisible $\cP\cT$-symmetric bilayer slab
whose optical potential is given by (\ref{v2=}) and
    \be
    u(x):=\left\{\begin{array}{cc}
    \fK^2[1-(\eta+i\kappa)^2]& {\rm for}~ x\in[-\ell/2,0),\\
    \fK^2[1-(\eta-i\kappa)^2]& {\rm for}~ x\in[0,\ell/2],\\
    0 &{\rm otherwise},\end{array}\right.
    \label{PT-u=}
    \ee
where $\eta$ and $\kappa$ are real numbers determining the
refractive index of the two layers as $\eta\pm i\kappa$, and
$\ell:=L_2/L$. Ref.~\cite{pra-2013} provides a detailed analysis of
the unidirectionally invisible configurations of  $\cP\cT$-symmetric
bilayer slabs. This allows for finding right-invisible configuration
with unit left reflection amplitude at desired wavelengths
$\lambda_0$. A simple example is a $\cP\cT$-symmetric bilayer slab
with
    \begin{align}
    &\eta = 2.996356,
    \quad\quad
    \kappa = 3.388790 \times10^{-3}, \\
    &L_2=\left\{\begin{array}{ccc}
        318.540\,\mu{\rm m}  &{\rm for}~\epsilon=0.05,\\[3pt]
        301.775 \,\mu{\rm m} &{\rm for}~\epsilon=0.10.
        \end{array}\right.
    \label{PT-L2=}
    \end{align}
Figure~\ref{fig5} provides a graphical demonstration of the
nonlinear amplification effect for the system depicted in
Fig.~\ref{fig1} when we identify the linear slab $\cS_2$ with the
$\cP\cT$-symmetric bilayer given by (\ref{v2=}) and (\ref{PT-u=}) --
(\ref{PT-L2=}). The physical quantities associated with the incident
wave and the Kerr slab are the same as those used to obtain
Fig.~\ref{fig2}.
Comparing Figs.~\ref{fig2} and \ref{fig5} we see that the choice of
the permittivity profile for the linear slab does not affect the
general behavior of the system. In particular away from the
resonance wavelength $\lambda_0$ it displays a strong filtering
effect.

A curious question regarding the  nonlinear amplification scheme we
have developed is weather it complies with the conservation of
energy. The scheme amplifies waves which escape to infinity,
therefore it should have a source of energy. Because the Kerr slab
has a real linear permittivity and Kerr coefficient, it cannot act
as an energy source. This suggests that the energy carried away by
the amplified wave is to be produced by the linear slab. This indeed
agrees with the presence of gain regions in the linear slab.
Therefore to maintain its function, we need to pump it with energy.
The system spends part of this energy to amplify the transmitted
wave. In this sense, the linear slab plays two important roles: 1)
It produces the necessary interference effect that eliminates the
need for injecting a high-intensity wave form $x=+\infty$ to realize
the near blow-up solution; 2) It produces the energy necessary for
amplifying the left-incident wave.

The presence of two slabs in our system raises the question whether
it is just a laser cavity with $k_\star$ being one of its lasing
modes. This is actually not true, because the amplification scheme
it operates upon is nonlinear, i.e., it only amplifies an incident
wave if it has the correct (and sizable) intensity. The basic
mathematical concept underlying the amplification effect associated
with laser cavities is that of a spectral singularity
\cite{prl-2009,pra-2011a}. This corresponds to the scattering
solutions of the linear Helmholtz equation that behave as zero-width
resonances. At a spectral singularity both the reflection and
transmission amplitudes of the system diverge. Because this happens
independently of the amplitude of the incident wave, the system can
amplify the background noise to sizable intensities and emit purely
outgoing coherent waves. The nonlinear amplification scheme we have
outlined in the present article makes use of a fundamentally
different mathematical phenomenon, namely the blow-up solutions of
nonlinear equations. This in particular implies that it cannot be
employed to amplify the background noise. It amplifies a
left-incident wave only if it has a particular (and generally large)
intensity. This in turn implies that one cannot operate the setup
for $k=k_\star$, because this would give rise to an infinite
amplification of an already high-intensity incident wave, which
would damage the system. It can only be operated for $k<k_\star$
where it would amplify the high-intensity incident wave to a much
larger intensity.

An important problem regarding the experimental realizations of our
nonlinear amplification scheme is the presence  of loses in
realistic Kerr slabs, which corresponds to situation where
$\widehat\varepsilon_l$ or $\sigma$ take complex values. This
obstructs the exact solvability of the corresponding nonlinear
Schr\"odinger equation \cite{marburger,yeh,chen}, but does not
affect the existence of blow-up solutions as long as the real part
of $\sigma$ is negative. Ref.~\cite{p152} establishes the existence
of blow-up solutions for the more general situations where
$\widehat\varepsilon_l$ and $\sigma$ are continuous complex-valued
functions of $z$ with the real part of $\sigma$ having a negative
upper bound, i.e., there is a real number $s_{\rm max}$ such that
$\RE[\sigma(z)]\leq s_{\rm max}<0$ for all $z\in[0,L]$. In
particular, the initial values $\sE(0)$ and $\sE'(0)$ determine a
blow-up solution of the Helmholtz equation for such a Kerr slab
provided that $\RE[\sE(0)^*\sE'(0)]>0$ and $L\geq 2.023\times\left\{
k^2|s_{\rm max}|\RE[\sE(0)^*\sE'(0)]\right\}^{-1/3}$, \cite{p152}.
These results provide the theoretical grounds for comprehensive
studies of more realistic applications of the nonlinear resonance
phenomenon we have introduced in this article.

\vspace{6pt} \noindent {\em {\bf Acknowledgements}:} We would like
to thank Kaan G\"uven for suggesting
Refs.~\cite{kaipurath,caspani,alam}, and Varga Kalantarov, Aref
Mostafazadeh, and Neslihan Oflaz for illuminating discussions. This
work has been supported by the Scientific and Technological Research
Council of Turkey (T\"UB\.{I}TAK) in the framework of the project
no: 114F357, and by the Turkish Academy of Sciences (T\"UBA).

\vspace{6pt} \noindent {\em {\bf Appendix.} A sufficient condition
for the presence of gain regions:} ~~Consider a possibly
complex-valued finite-range potential $v(x)$. Using an appropriate
translation and dilation of the independent variable, we can
identify the support of this potential with the unit interval,
$[0,1]$, i.e., without loss of generality, we suppose that $[0,1]$
is the smallest closed interval outside of which $v(x)$ vanishes. We
can use $v(x)$ to describe the interaction of a normally incident TE
wave with a planar slab placed in vacuum. Suppose that the slab lies
between the planes $\rz=0$ and $\rz=a$ and has a refractive index
$\fn(\rz)$. Then, $v(x)$ is the optical potential for the slab
provided that
    \be
    \fn(\rz)=\sqrt{1-\frac{v(\rz/a)}{\fK^2}}.
    \label{n-app}
    \ee
Here $\fK=ka$ and $k$ is the wavenumber of the incident wave
\cite{silfvast}. The Helmholtz equation for this system is
equivalent to the Schr\"odinger equation,
    \be
    -\psi''+v(x)\psi(x)=k^2 \psi(x).
    \label{sch-eq-app}
    \ee

Because imaginary part of $\fn(\rz)$ is typically much smaller in
magnitude than its real part, the regions in which it takes negative
values coincide with those where imaginary part of $v(x)$ is
positive; $\IM[\fn(\rz)] $ and $\IM[v(\rz/a)]$ have opposite sign.
The regions in which $\IM[\fn(\rz)]<0$ are called the gain regions,
because the propagating waves are amplified while passing through
them \cite{silfvast}. We therefore call a region $\sG$ of the real
axis ``a gain region,'' if $\IM[v(x)]>0$ for all $x\in\sG$.
Similarly, a lossy region $\sL$ is defined by the condition:
$\IM[v(x)]<0$ for all $x\in\sL$. \vspace{3pt}

\noindent {\em Theorem:} Let $v(x)$ be a finite-range potential with
its left reflection and transmission amplitudes, $R^l$ and $T$,
satisfying
    \be
    |R^l|^2+|T|^2>1.
    \label{condi-1}
    \ee
Then the support of $v(x)$ must include gain regions.\vspace{3pt}

\noindent {\em Proof:} First, we multiply both sides of
(\ref{sch-eq-app}) by $\psi(x)^*$ and write the result as
    \[[\psi(x)^*\psi'(x)]'=|\psi'(x)|^2+[v(x)-k^2]|\psi(x)|^2.\]
Evaluating the imaginary part of the left-hand side of this equation
and integrating it over the support of $v(x)$, which we identify
with $[0,1]$, we find
    \be
    \IM[\psi(x)^*\psi(x)']\Big|_0^1=\int_0^1 dx\: \IM[v(x)]\:|\psi(x)|^2.
    \label{app-eq-1}
    \ee
Now, consider the case that $\psi(x)$ is a scattering solution of
(\ref{sch-eq-app}) corresponding to a left-incident wave, i.e.,
    \be
    \psi(x)=\left\{
    \begin{array}{ccc}
    A_-[e^{ikx}+R^le^{-ikx}] & {\rm for} & x\leq 0,\\
    A_-T e^{ikx} & {\rm for} & x\geq 1.
    \end{array}\right.\nn
    \ee
Substituting this equation in (\ref{app-eq-1}) yields
    \be
    |R^l|^2+|T|^2-1=\frac{1}{k|A_-|^2}\int_0^1 dx\: \IM[v(x)]\:|\psi(x)|^2.
    \ee
If the support of $v(x)$ has no gain regions, $\IM[v(x)]\leq 0$ for
all $x\in[0,1]$, and the right-hand side of this equation cannot
take a positive value. But according to (\ref{condi-1}), its
left-hand side is positive. This implies the presence of gain
regions.~~~$\square$ \vspace{3pt}

It is easy to see that the statement of this theorem also holds for
the scattering potentials \cite{bookchapter} having an infinite
range.

\ed

\ed
\begin{thebibliography}{99}
\bibitem{thm2} This statement holds also for $\hat\varepsilon(x,y)=\hat\varepsilon_b[1+e^{2i\alpha y}u(x,u)]$ with $\hat\varepsilon_b>0$ provided that we replace $k$ by $\sqrt{\hat\varepsilon_b}\,k$.
\bibitem{footnote1} This follows from the fact that in this case we can expand the right-hand side of (\ref{TM-v}) in a convergent power series involving products of powers and derivative of $\eta(x,y)$ and the fact that their Fourier transform with respect to $y$ has the same property as $\tilde\eta(x,\fK_y)$.
\end{thebibliography}
